\documentclass[manuscript,nonacm]{acmart}

\usepackage{makecell}
\usepackage{array}
\usepackage{multirow}

\AtBeginDocument{%
  \providecommand\BibTeX{{%
    \normalfont B\kern-0.5em{\scshape i\kern-0.25em b}\kern-0.8em\TeX}}}

\setcopyright{acmcopyright}
\copyrightyear{2018}
\acmYear{2018}
\acmDOI{XXXXXXX.XXXXXXX}

\acmConference[Conference acronym 'XX]{Make sure to enter the correct
  conference title from your rights confirmation emai}{June 03--05,
  2018}{Woodstock, NY}
\acmBooktitle{Woodstock '18: ACM Symposium on Neural Gaze Detection,
 June 03--05, 2018, Woodstock, NY} 
\acmPrice{15.00}
\acmISBN{978-1-4503-XXXX-X/18/06}

\begin{document}

\title{An Interaction Design Toolkit for Physical Task Guidance with Artificial Intelligence and Mixed Reality}

\author{Arthur Caetano}
\email{caetano@ucsb.edu}
\orcid{0000-0003-0207-5471}
\affiliation{%
  \institution{University of California}
  \city{Santa Barbara}
  \state{CA}
  \country{USA}
}

\author{Alejandro Aponte}
\email{aponte@ucsb.edu}
\orcid{0000-0003-2122-0451}
\affiliation{%
  \institution{University of California}
  \city{Santa Barbara}
  \state{CA}
  \country{USA}
}

\author{Misha Sra}
\email{sra@ucsb.edu}
\orcid{0000-0001-8154-8518}
\affiliation{%
  \institution{University of California}
  \city{Santa Barbara}
  \state{CA}
  \country{USA}
}

\renewcommand{\shortauthors}{Caetano et al.}


\begin{abstract}
Physical skill acquisition, from sports techniques to surgical procedures, requires instruction and feedback. In the absence of a human expert, Physical Task Guidance (PTG) systems can offer a promising alternative. These systems integrate Artificial Intelligence (AI) and Mixed Reality (MR) to provide realtime feedback and guidance as users practice and learn skills using physical tools and objects. However, designing PTG systems presents challenges beyond engineering complexities. The intricate interplay between users, AI, MR interfaces, and the physical environment creates unique interaction design hurdles. To address these challenges, we present an interaction design toolkit derived from our analysis of PTG prototypes developed by eight student teams during a 10-week-long graduate course. The toolkit comprises Design Considerations, Design Patterns, and an Interaction Canvas. Our evaluation suggests that the toolkit can serve as a valuable resource for practitioners designing PTG systems and researchers developing new tools for human-AI interaction design.

\end{abstract}



\received{20 February 2007}
\received[revised]{12 March 2009}
\received[accepted]{5 June 2009}

\maketitle
\section{Introduction}
Mixed reality (MR) technology can display interactive virtual content anchored to the real environment~\cite{milgram1995augmented}. This capability is suitable to offer instructions to tasks situated in the real environment~\cite{curtis1999several}. Such instructions can be contextualized to users' needs using artificial intelligence (AI), with examples including early instances of AI such as rule engines~\cite{feiner1993knowledge} and more recent developments such as large language models (LLMs)~\cite{wu2024artist}. AI has also been used to offer personalized feedback in real world tasks through MR interfaces, leveraging sensing and error detection techniques~\cite{anderson2013youmove}. In this work, we refer to the class of MR systems that display AI-generated instruction and feedback in a task situated in the real environment as \textbf{Mix}ed Reality \textbf{I}ntelligent \textbf{T}ask \textbf{S}upport (MixITS). MixITS systems have been proposed for several task domains including cooking~\cite{sosnowski2023challenges, wu2024artist}, machine maintenance~\cite{feiner1993knowledge}, and fitness training~\cite{anderson2013youmove, mandic2023arfit}.

MixITS have the potential to improve user performance and facilitate physical skill acquisition by overcoming limitations of other forms of instruction and feedback. Textual and audiovisual formats, such as books and video-tutorials, lack personalization and proactive interventions to prevent or correct user errors, all of which can be achieved with MixITS. Human coaches can benefit from the deployment of MixITS systems as they allow tracking and realtime analysis of multiple physiological and kinematic signals including full body pose, eye-gaze, and heart rate~\cite{bernal2022galea} which could go unnoticed by humans. In the absence of a human expert due to schedule, budget or geographical constraints, MixITS can act as an automated coach to ensure consistent training and performance of end-users.

MixITS could enable a large audience to acquire new physical skills and perform tasks in the real world in a safe and precise manner. However, research and design of MixITS can be challenging due to the inherited complexity of AI and MR technologies. Previous research has identified challenges that AI introduces to HCI~\cite{yang2020re}, emphasizing the need for new design guidelines and toolkits specific to AI-based applications. Separate studies have pointed factors of MR that contribute to complexity in design, such as understanding of users’ surroundings and relationships between virtual and real elements~\cite{freitas2020systematic}. Although the existing AI~\cite{amershi2019guidelines, yang2020re, yildirim2022experienced, feng2023ux} and MR~\cite{laviola2017ui, ultraleap2024guide, meta2024guide, apple2024guide, microsoft2024guide, design2024double} provide a useful starting point, considerably less research has been done on design aids at the intersection of these technologies~\cite{xu2023xair}.

In addition to the challenges inherited from AI and MR, the application domain of task support situated in the real environment adds another layer of complexity. Examples of critical design decisions in MixITS include: How much guidance should be provided to balance correct execution and learning? Should the system always intervene in case of mistake, or should it balance the negative impact on the outcome and keep the user focused on the task? How to promote user trust on the system given the AI can make mistakes and provide instructions that diverge from the users' embodied knowledge of the task? Limited research has been done in eliciting MixITS design challenges and developing structured design tools in this domain are non-existent.

To fulfill this gap, we propose MixITS-Kit, a structured set of design tools to tackle the complexity of MixITS at multiple abstraction levels, contributing to faster developments in this application domain. Our toolkit is comprised of three elements:

\begin{enumerate}
    \item \textbf{Interaction Canvas}: a visual tool to streamline the analysis of interactions between the users, system, and the environment (Section~\ref{sec:canvas}). 
    \item \textbf{Design Considerations}: a catalog of high-level design considerations in MixITS (Section~\ref{sec:themes}).
    \item \textbf{Design Patterns}: a set of lower level describing common problems in MixITS, observed solutions, and application examples to guide prototyping and development (Section~\ref{sec:patterns}).
\end{enumerate}

We created MixITS-Kit based on observation and documentation of eight low-fidelity MixITS prototypes developed by 25 graduate students during a graduate 10-week human-AI interaction (HAI) class. We used Don Norman's Gulfs of Execution and Evaluation model as the foundation for our Interaction canvas that allows designers to analyze potential interaction problems between the actors involved in a MixITS scenario. The Interaction Canvas also enabled us to systematically identify problems and corresponding design solutions in the prototypes, facilitating design pattern elicitation~\cite{winters2009dealing, retalis2006eliciting, borchers2000pattern}. Through a reflexive thematic analysis~\cite{braun2006using, braun2024deductive} of the projects, we distilled six overarching design considerations for MixITS systems.

To evaluate the benefits and limitations of our toolkit, we conducted a user study where a separate group of nine participants used MixITS-Kit to solve a series of design problems. This evaluation assessed ease of use, toolkit research goals proposed in prior literature~\cite{ledo2018evaluation}, and potential use as shared vocabulary. Our results demonstrate that participants were able to utilize MixITS-Kit to propose multiple solutions for common challenges in MixITS design. User feedback also points to future work in an interactive version of our toolkit, especially to streamline gulf analysis and navigation in the design pattern catalog.

MixITS-Kit offers a structured set of tools that fulfill gaps in existing design aids that consider only AI or MR alone and tackle the unique challenges of task support situated in the real environment from different levels of abstraction. MixITS-Kit contributes to consolidating and disseminating design practices in MixITS design, favoring faster development of better systems that can help a broader audience to learn physical skills or perform real-world tasks in a safe and efficient manner.

\section{Related Work} \label{sec:rw}
In this section, we review literature on MR task support, which is similar to MixITS but lacks an AI component. We then focus on prior research in MixITS systems, which addresses technical challenges in underlying technologies but provides limited tools for interaction design. Finally, we discuss methods used in earlier work to derive design guidelines for human-AI interaction and mixed reality design. While these tools are useful for MixITS design, they overlook many unique problems of this domain, which we address in later sections.

\subsection{Mixed Reality Task Support}

Supporting users in tasks situated in the real environment has been a long lasting motivation for MR research. A 1990 pioneer study demonstrated the feasibility of using MR to instruct workers in an aircraft wiring assembly task~\cite{thomas1992augmented, curtis1999several}. However, evaluations showed no significant reduction in task completion time. Research associated this result with the need for improved instructional design, ergonomic issues and social acceptance of the equipment---concerns that remain relevant today.

Later studies compared MR instructions with alternatives in several real environment tasks. In a warehouse pick-up task, researchers demonstrated improvements in task performance when using MR instructions instead of paper and audio baselines~\cite{weaver2010empirical, schwerdtfeger2006mobile}. In a manual assembly task, an MR instruction condition over performed paper and 2D displays in terms of error rate and mental effort~\cite{tang2003comparative}. In a machinery repair task, MR instructions allowed mechanics to locate tasks more quickly and with less head movement than when using a 2D display~\cite{henderson2010exploring}. ARDW designed and evaluated an MR system to support printed circuit board testing~\cite{chatterjee2022ardw} and reported reduced context-switching between instruction and the task, significantly faster localization on the board, and highlighted the need for fine alignment between virtual elements and physical elements.

Tiator et al.~\cite{tiator2018venga} proposed an MR instruction system to assist climbers, highlighting benefits such as maintaining appropriate challenge levels and documenting ascents. In Skillab~\cite{shahus2023skillab}, researchers combined MR instructions with haptic assistance through electrical muscle stimulation in a floor lamination task. Their system offered better user experience and outcome quality than a paper-based baseline. Their prototype identified the need to allow users to modify specific steps according to their preferences and prior knowledge. This work demonstrates the potential of combining MR instruction with physical actuators and the need for adaptability. MR has been used to support piano learning with positive effects on learners' motivation, ability to read music scores, and play notes correctly~\cite{rigby2020piarno}. In the same study, researchers emphasize the need of offering customization options to meet users' preferences and skill level.

These relevant studies and systems offer evidence of the benefits of MR instruction for tasks situated in the real environment and contribute to design considerations that resonate with the ones derived from our own study. However, they do not include the AI component present in the MixITS definition. Our work corroborates previous findings, such as the need to adapt to users' preferences and skill level, and potential to incorporate physical actuators and additional sensors missing in current commercial MR devices. Differently from prior work, MixITS-Kit adds to their design considerations by accounting for the presence of AI in the design.

\subsection{Mixed Reality Intelligent Task Support}
\label{sec:rw_mixit}
More directly related to our work, a pioneer system proposed by Feiner et al~\cite{feiner1993knowledge} presented a augmented reality (AR) interface combined with a knowledge-base to automate the design of instructions for maintenance and repair tasks. This is one of the earliest examples of what we refer to in this paper as MixITS. Their work identified potential for multimodal instructions and suggested MixITS will become the preferable way to learn complex real environment tasks in the future. More recent work in MixITS include models of user-gaze behavior too estimate visual attention and expertise level in cooking and coffee brewing tasks~\cite{yoo2023modeling}. ARTiST~\cite{wu2024artist} used LLM few-shot prompting to optimize the length and the semantic content of textual instructions displayed in MR for two tasks---cooking and spatial localization. This intervention alleviated participants' cognitive load and improved task performance when compared with unprocessed instructions. Also in a cooking task, ~\cite{rheault2024predictive} proposed a MixITS that leverages action recognition, error and object detection to intelligently provide instruction and feedback.

To facilitate the design and evaluation of MixITS systems, prior work has also proposed technical platforms including open source MR systems combined with vision and language models~\cite{bohus2024sigma} and a visual analytics system~\cite{castelo2023argus}. Although these platforms alleviate the engineering and data analytics burden, they offer limited guidance to the user-centered design of novel MixITS systems. Our work aims to address this gap, providing a toolkit that is specifically designed for MixITS systems. By doing so, we hope to encourage the development of more diverse MixITS applications. Currently, very few such systems exist, limiting our ability to evaluate them at scale. As more MixITS systems are built and deployed, we will be better positioned to identify and understand new real-world challenges as they emerge, furthering our knowledge and improving future designs in this critical area of human-AI interaction.

\subsection{Designing with AI}

Researchers have developed human-AI interaction guidelines and explored the role of designers in AI teams through design workshops, product reviews, and literature analyses. These studies have also identified design challenges stemming from AI's uncertain capabilities and complex outputs.

For example, Amershi et al.~\cite{amershi2019guidelines} developed 18 human-AI interaction design guidelines, derived from over 150 recommendations from academic and industry sources, and validated through evaluations including HCI practitioner testing. The authors noted the need for specialized tools to address unique design challenges in AI-integrated interfaces, particularly in modeling interactions involving the physical context. A follow-up study presents an evaluation protocol for human-AI interaction, focusing on productivity applications such as document and slide editors, search engines, email, and spreadsheet applications \cite{li2023assessing}.

Delving into AI system design, Yang et al.~\cite{yang2020re} identified two key complexities that designers face: uncertainty about AI capabilities, and varying complexity of the types of output AI models generate. They proposed a four-level classification, with level-four systems being the most challenging to design due to continuous learning and adaptive outputs. MixITS systems fall into this category. The authors suggest that existing design methods are inadequate to address the range of potential AI behaviors in real-world contexts for these complex systems.

Exploring how experienced designers use AI in enterprise applications, Yildirim et al.~\cite{yildirim2022experienced}, found AI was used in both UI design and higher-level systems and services. They identified tools such as data-annotated service blueprints and wireframes that designers use to work with AI. The study highlights the need for AI-specific design tools, arguing that traditional UX practices require adaptation to address the unique challenges and opportunities presented by AI technologies.

Expanding on previous studies, Feng et al. ~\cite{feng2023ux} conducted a contextual inquiry with 27 industry UX practitioners to explore challenges in AI integration. They identified specific issues UX practitioners face when integrating AI into designs. These include communicating AI capabilities, calibrating user trust, addressing explainability, and merging data science with user-centric design. Their work introduced concepts like ``AI model fidelity'' and ``probabilistic user flows'' to aid designers in AI application development. They also noted the importance of involving domain experts in collaborative human and AI design approaches, even if they lack UX or AI expertise. These tools are particularly relevant for MixITS systems, where AI behavior uncertainty significantly impacts users.

While tools and guidelines from prior work provide valuable insights for general AI interaction design, they overlook many of the unique MixITS challenges derived from the combination of MR and tasks situated in the real environment. For example, AI understanding of the physical world might be misaligned with the users' embodied knowledge, resulting in incorrect instructions. Another design problem not addressed by current HAI guidelines is timeless and adequate modality for AI interactions, that might ignore the real environment task at hand is priority for the user. Informed by prior HAI research and careful analysis of eight MixITS prototypes, our MixITS-Kit offers specialized design tools for intelligent task support in mixed reality.

\subsection{Designing with MR}

Challenges of designing with MR described in literature~\cite{ashtari2020creating} include difficulty to anticipate users' movements in the real environment and dealing with distractions from the real environment. Other studies highlighted the difficulties of aligning physical and virtual elements spatially and semantically~\cite{ellenberg2023spatiality}. To better support MR designers, industry and academia have proposed design tools.

Industry practitioners have compiled valuable guidelines for designing MR experiences. Guidelines such as ``design to avoid occlusion'' and ``design for the interaction zone'' can help designers to create more usable and comfortable interactions in MR considering tracking limitations and user upper body reach limits~\cite{ultraleap2024guide}. Recommendations range from spatial layout to interaction such as ``size and distance for proper depth perception'' and ``give users (...) multi-modal input, such as hand ray-and-speech input (...)'' to help guide the design of MR experiences~\cite{meta2024guide}. Best practices suggest displaying content within the user's field of view, supporting ``indirect gestures'' (i.e., gestures executed in resting position), and avoiding the display of overwhelming motion to ``prioritize comfort''~\cite{apple2024guide}. Industry has also proposed design processes and techniques for designing MR applications. Low fidelity prototyping techniques such as ``bodystorming,''~\cite{burns1994actors, oulasvirta2003understanding}, i.e., manipulating low-cost, tangible props that represent components of a MR application, are a useful method to validate and refine early design concepts~\cite{microsoft2024guide}. Acting out a scenario is recommended to gain perspective on how a user would interact with an MR application~\cite{microsoft2024guide}. Additional guidelines for frequent tests and design iterations to better understand user behavior are also suggested~\cite{design2024double, meta2024guide}. We incorporated these techniques into the assignments for our design course to help students prototype MixITS systems without feeling constrained by current technological limitations or implementation challenges.

\cite{laviola2017ui}, proposed practical guidelines for designing 3D UIs---including MR interfaces---such as ergonomics and comfort (e.g., ``Design for comfortable poses''), user safety (e.g., ``Provide physical and virtual barriers to keep the user and the equipment safe''), and interaction design (e.g., ``Consider using props and passive feedback, particularly in highly specialized tasks''). A survey on human remote collaboration through MR reviewed extensive research and identified design choices in local interfaces that can be helpful in MixITS systems as well~\cite{wang2021ar}. The same study presents a comprehensive list of technological toolkits, but does not point to design toolkits. The domain of human remote collaboration holds similarities to MixITS but the substitution of the remote expert by and AI creates unique challenges, for example in modulating instruction and feedback frequency and handling false positive error detection, motivating our research on a specialized design toolkit, MixITS-Kit.

Previous research has identified design patterns through a structured reflection on artifacts created during the iterative development of a MR industrial safety training system~\cite{rauh2024navigating}. This approach aligns closely with our methodology, as both rely on post-project reflection on project documentation to extract design patterns relevant to the MixITS domain. However, their study does not address the unique potentials and challenges introduced by incorporating AI-driven instruction and feedback.

Valuable studies have devised guidelines at the intersection of MR and AI, which is closely related to MixITS-Kit haven't concentrated the use case of real environment task support. XAIR~\cite{xu2023xair} is a design toolkit for explainable AI in augmented reality (AR). It was developed through a multidisciplinary literature review, surveys, workshops, and user studies. It offers guidelines for determining when, what, and how to explain AI outputs to AR users. While XAIR provides valuable insights for integrating explainability into MixITS systems, addressing factors like user cognitive states and environmental context, our toolkit takes a broader approach to MixITS system design challenges beyond explainability alone.

Following a review of 311 papers covering XR and AI, published between 2017 and 2021, Hirzle et al.~\cite{hirzle2023xr} identified research opportunities in the intersection of XR and AI. Our work builds on Hirzle et al.'s recommendations for future research by providing design tools to analyze human-AI interaction challenges in XR, specifically in the context of physical task guidance.

\section{Method}

We aim to offer designers a structured toolkit for MixITS design that helps to tackle interaction problems between the user, system, and the real environment from multiple levels of abstraction. Prior research has leveraged widely available products in well-documented domains such as web search and recommendation systems to iteratively derive human-AI guidelines~\cite{amershi2019guidelines}. Unfortunately, this method is not directly applicable to the MixITS domain at the moment, due to the scarcity of documentation focusing on the design and the general availability of products in this space, as presented in Section~\ref{sec:rw_mixit}. Another obstacle to the development of MixITS design tools is the engineering complexity, long development cycles, and high costs associated with integrating AI into real-time MR applications.

To overcome the lack of availability of existing MixITS systems their development costs, we adopted a middle-ground approach using low-fidelity prototypes created by graduate students in a 10-week long course. We employed techniques like bodystorming~\cite{burns1994actors, oulasvirta2003understanding}, video prototyping~\cite{leiva2020pronto}, and role-playing~\cite{svanaes2004putting} to explore potential design challenges without full implementation. This method yielded valuable insights, leading to high-level design considerations and low-level solutions for identified problems presented as design patterns.  Furthermore, the focus on problem definition and low-fi prototyping allowed us to propose design tools that are agnostic to the current technological limitations. We base our analysis methodology on prior research on design with AI~\cite{lupetti2024making} while addressing common challenges associated with using design workshops in research~\cite{elsden2020design}.

We structured the design course to address common challenges associated with using design workshops in research~\cite{elsden2020design}. By coupling in-class discussions and multiple prototype iterations, students engaged deeply with the material, resulting in higher-quality outputs than would be possible in shorter, broader-audience workshops. Students had dual motivations---enhancing their knowledge of human-AI interaction and MR while meeting academic requirements---which contributed to sustained engagement and meeting objectives over the 10 weeks. The structured assignment schedule with in-class presentations and feedback not only facilitated a continuous data collection process but also enabled systematic organization and allowed for deeper analysis of the evolving design prototypes, reflecting the iterative nature of the design process.

Class activities and assignments were designed with the pedagogical goal of educating future AI researchers about challenges in MixITS system design and exploring potential solutions. The curriculum aimed to help students understand the problem domain of human-AI interaction in MixITS systems, as well as the challenges and complexities involved in developing realtime, AI-supported interactive systems for task assistance. Our local IRB approved the use of the assignment deliverables as research data. Students provided informed consent for their assignments to be used for research purposes after the academic term ended and grades were submitted. One team of four members did not consent to their assignment data being used for research so we excluded their data from our analysis.

\subsection{Participants}
The course had 29 graduate students from computer science, engineering, and neuroscience at our university, divided into nine teams — seven teams of three members and two teams of four members. Their diverse backgrounds produced teams with a balanced understanding of AI, MR, and human factors. All students were new to the MixITS application domain. We consider our student pool to be representative of early-career designers, engineers, and researchers new to the MixITS domain but with prior AI, XR, and HCI experience acquired through courses or research projects.

\subsection{Apparatus}

\subsubsection{Syllabus and Readings}
The course syllabus included a variety of learning activities designed to provide a comprehensive understanding of the different aspects of designing MixITS systems. The activities included lectures on topics such as introduction to AI and HCI, design thinking, human-centered design, low fidelity prototyping techniques, AI interpretability and explainability, human-AI interaction, and ethics. In addition to lectures, the course included frequent in-class discussions on selected readings. Students wrote weekly reflections on assigned readings, responding to three specific questions about each reading. Students also commented on their classmate's reflections. The course included project-related checkpoints where students presented their progress in class and received feedback from their peers and the instruction team. Assignments were designed to align with the lectures and readings to encourage students to apply the material to their practical design challenges. A critical component of the class was interactive activities earlier in the quarter, including role-playing exercises with props, where students acted out physical tasks (e.g., making coffee), played the roles of an AI agent and an interface, to gain a hands-on understanding of the challenges and considerations involved in designing MixITS systems.

\subsubsection{Role-played MixITS Scenario}\label{sec:roleplay}
The in-class coffee-making demonstration and role-play scenario highlighted key challenges in AI-based MixITS systems, providing students with a hands-on understanding of the complexities involved in real-time AI instruction and feedback. Several limitations of MixITS systems became apparent, including action misinterpretation and inability to handle unexpected events not present in the training data. The system's failure to detect user errors, such as adding salt instead of sugar, was also highlighted. Timing and sequence issues in managing procedural steps, along with environmental variables affecting object and action recognition, further presented aspects to consider in developing MixITS systems. The demonstration revealed how user expertise levels and mental models can significantly influence their interactions with AI systems. Novices and experts may approach and respond to AI guidance differently based on their understanding and expectations of the system's capabilities. By confronting realistic AI limitations, students were better prepared to design realistic and robust MixITS systems in their projects.

\subsubsection{Group Project and Assignments}
The primary outcome of the course was a group project where students designed and created low-fidelity prototypes of their MixITS systems. The project was broken down into five bi-weekly assignments. Students used tools such as empathy maps, sketches, paper, and video prototypes of role-played scenarios for their design assignments, which they presented in class for feedback. They were expected to integrate the feedback into subsequent iterations of their prototype. Written assignments for each prototype stage involved design reflections, anticipated AI and user errors with potential solutions, interface design elements, user evaluation processes and outcomes, and more.

\begin{figure}
    \centering\includegraphics[width=1\linewidth]{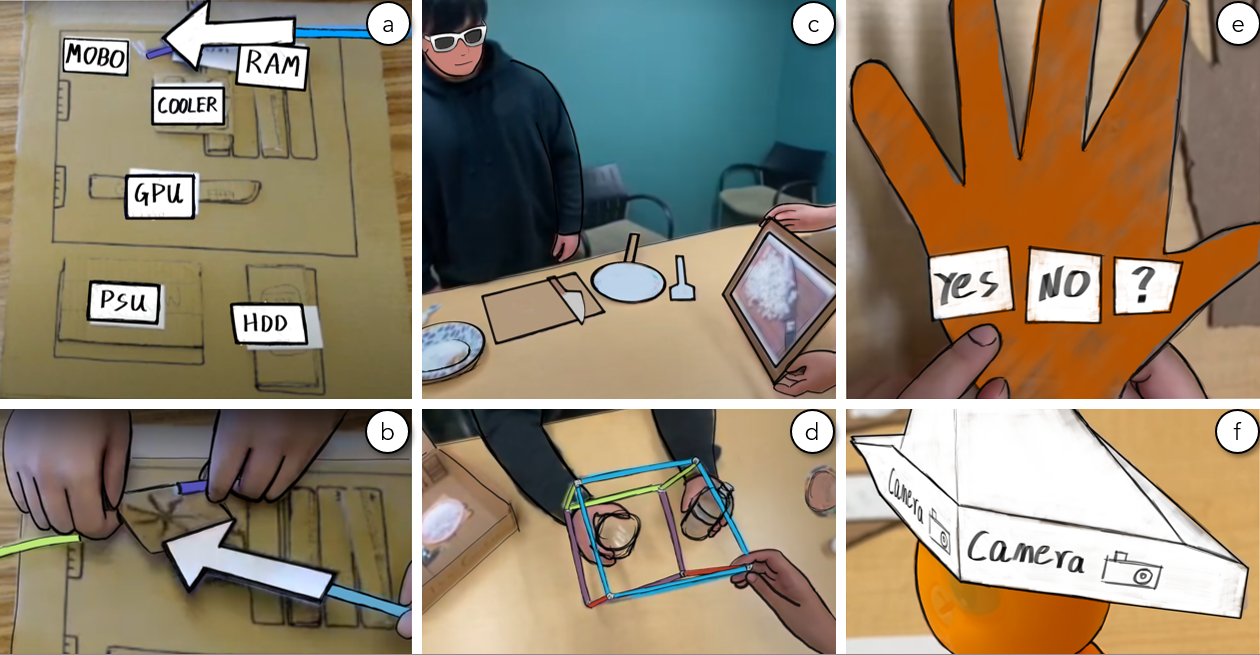}
    \caption{
    Low-fidelity prototyping allowed students to iterate on the design of MixITS systems with a small budget. Students created cardboard props representing real objects, such as a PC motherboard \textbf{(a)} for AI-assisted repair and kitchen utensils \textbf{(c)} for AI-assisted cooking. The MR interface prototypes used low-cost materials like paper labels to simulate virtual labels on a PC motherboard \textbf{(a)}. Students moved these props to prototype dynamic behaviors, such as a virtual arrow pointing to parts of a PC \textbf{(b)} and tracking ingredients with bounding boxes made of straws\textbf{(d)}, similar to low-fidelity paper prototyping of mobile or web applications. They also used inexpensive materials to represent customized hardware, including a haptic feedback glove \textbf{(e)} and a 360-degree camera headset \textbf{(f)}. To simulate large outdoor environments, such as a four-way stop for an AI-supported navigation app for blind users, students created miniature versions using dioramas, figurines, and toys with extra sensors such as cameras embedded in paper hats. This approach allowed them to enact full system functionality in a controlled, scaled-down setting that represented complex real-world scenarios\textbf{(f)}.}
    \label{fig:class}
\end{figure}

\subsection{Procedure}
\label{sec:class_procedure}
Our human-AI interaction course focused on MixITS scenarios was 10-week long with lectures of one hour and 15 minutes twice a week. The syllabus included MR design, human-centered AI, and prototyping techniques suitable. Throughout the course, each team developed a low-fidelity MixITS system with an incremental and iterative process. At the end of the course they evaluated their prototypes with users. Students documented the entire process with weekly assignments, in-class presentations, written reports and recordings of user testing. This rich documentation was the empirical basis to develop the MixITS-Kit.

\subsubsection{Assignment 1 (Explore)}

In Week 2, Assignment 1, students analyzed a freely chosen physical task using flowcharts and sketches that detailed the task steps, required tools, expected outcomes, and surrounding environment. Next, they created empathy maps to gain a deeper understanding of their potential audience, adopting a user-centered approach. Students then reflected on the challenges, benefits, and risks of deploying an AI system to support users in the physical task. Their reflections resulted in AI-based use cases to facilitate learning and improve users' skills along with suitable interaction modalities. This assignment laid down the foundations for the next iterations.

\subsubsection{Assignment 2 (Empathize)}

The second assignment focused on two primary objectives: fostering empathy for potential users and deepening understanding of a MixITS task from both user and AI perspectives. This approach aimed to cultivate a more empathetic and comprehensive design process, considering the viewpoints of both the users and the AI system. Students participated in a role-play exercise to simulate the interactions between a user, an AI system, and the interface in a MixITS scenario (Section~\ref{sec:roleplay}). This activity incorporated established HCI methods such as the think-aloud protocol~\cite{ericsson1980verbal} and bodystorming~\cite{burns1994actors, oulasvirta2003understanding}. Through this hands-on experience, students gained insights into the complexities of human-AI interaction in a MixITS system, including the reality of AI errors, and identified interface design challenges to address these errors. Understanding the interface design challenges focused on managing AI errors while balancing several key factors: preserving user agency, minimizing cognitive load, reducing user frustration, maintaining trust in the system, and preventing users from disregarding the AI's input altogether. This approach aimed to help students recognize the various elements that need to be considered while designing human-AI interfaces that effectively support users without compromising the overall user experience or the system's utility.
For their submission, students recorded a prototype walkthrough and provided a written reflection on their prototype. This combination of practical demonstration and analytical reflection allowed students to synthesize their learning and apply it to their design scenario.

\subsubsection{Assignment 3 (Prototype)}

Assignment 3 built upon the insights gained from the previous two assignments, challenging students to prototype solutions for their MixITS scenario. Rather than designing entire systems, students narrowed their focus to critical aspects such as managing AI errors, addressing user perception and trust, and examining system design assumptions. Students created low-fidelity prototypes of MR interfaces, employing a Wizard of Oz technique~\cite{kelley1983empirical} to produce video prototypes. This approach allowed them to explore design concepts without committing prematurely to complex and time-consuming engineering, as pointed in prior work on MR prototyping~\cite{de2012mobile}. They detailed the physical tasks in their scenarios by visually mapping out step pre-conditions and task sequences, using the provided drip coffee-making scenario as a guideline. Through this assignment, students learned to appreciate the nuances of accommodating task variations and error detection in AI-assisted systems. They gained experience in balancing user needs with system capabilities, and understood the importance of considering execution uncertainties in their designs. This exercise reinforced the complexity of creating intuitive, effective, and trustworthy human-AI interactions in MixITS scenarios.

\subsubsection{Assignment 4 (Iterate)}

Assignment 4 focused on the evaluation of the refined prototypes, which now incorporated a simulated AI backend. The assessment process involved instructors acting as users to test both the interface's effectiveness and its technical design. To challenge the system's robustness, instructors intentionally introduced errors during testing when instructions were unclear or exposed to safety risks (e.g., cooking burns), helping students uncover design challenges relevant to real-world applications of MixITS systems. Students were tasked with detailing their user study design including tasks and metrics, writing reflective essays on their findings, itemizing instructor feedback and brainstorming potential solutions to observed interface failures. This reflective exercise served to deepen their understanding of their designs and provided valuable insights for future iterations of the projects.

\subsubsection{Final Assignment (Reflect)}

The final design project culminated in a comprehensive report on the designed MixITS system. This report discussed the rationale behind their designs, explored user benefits for learning new skills or performing tasks, and offered a critical reflection on the system's strengths and weaknesses. Students also provided recommendations for future designers working on similar projects. This final assignment encapsulated the entire design journey, integrating insights from previous assignments and incorporating feedback from peers and instructors. The project served as a reflection of their growth and understanding in designing AI-assisted MixITS systems.

\section{Data Analysis}
We analyzed all course data, including video prototypes and accompanying written documentation produced by each team for every assignment. We performed two types of analyses: 1) reflexive thematic analysis \cite{braun2006using} which resulted in six design considerations, and 2) design pattern elicitation ~\cite{winters2009dealing, retalis2006eliciting, borchers2000pattern} which resulted in the identification of 36 potential problems and example solutions. 

\subsection{Reflexive Thematic Analysis}
We used the flexible qualitative approach of reflexive thematic analysis, as formalized by Braun \& Clarke \cite{braun2006using}.
This method allows either inductive or deductive strategies~\cite{braun2024deductive}. We chose a deductive approach starting from goal-oriented codes taken from the metacognitive questions in the assignments. Braun \& Clarke emphasize that reflexive thematic analysis can be effectively conducted by a single analyst, and that inter-rater reliability is not necessary for the method to be applied rigorously~\cite{braun2006using, braun2024deductive, maxwell2010using}.

Our analysis followed the 6-stage procedure proposed by Braun \& Clarke \cite{braun2006using, braun2024deductive}. First, the analysis was conducted by one of the authors who did not have any classroom role or interaction with the student teams. This data analyst started by familiarizing themselves with the assignments. The initial set of codes was based on the metacognitive~\cite{flavell1976metacognitive} components in the assignments---design challenges, decisions, trade-offs, and recommendations for future designers. 
Following the initial coding, the analyst developed a set of themes. These were presented and discussed with the teaching team, to ensure they were coherent to the classroom experience and captured insights gained during the design course. The discussion resulted in a final set of themes that are named and reported in Section~\ref{sec:themes} as design considerations. 

\subsection{Design Pattern Elicitation} \label{subsec:designpattern}

Drawing inspiration from previous work on design pattern elicitation ~\cite{winters2009dealing, retalis2006eliciting, borchers2000pattern}, our method to elicit MixITS design patterns consisted of, (1) compiling case studies, (2) identifying common functionalities, (3) decomposing the functionalities into triplets of \textit{context, problem, and solution,} and (4) refining the triplets by detailing, merging, splitting, and removing.
The eight prototypes were used as case studies, represented by written reports and video recordings. We reviewed the reports and watched the videos for each prototype, to compile a coarse list of 63 functionalities identified across the eight prototypes. Each of the 63 functionalities was broken down into triplets of \textit{context, problem, and solution.} Following that, the 63 triplets were labeled with one of the eight Interaction Gulfs presented in Section~\ref{sec:gulfs}. The Gulf labels abstract the details of the context and for this reason the full context description was omitted from Table~\ref{tab:full}. The gulf labeling step involved identifying the actor (Human or AI) and the target (Human, AI, or the Environment) of the interaction. Additionally, the actor's goal and the target's means to enable the user to accomplish their goals, in the execution cycle, and the actor's interpretation of the target's feedback, in the evaluation cycle, were labeled. Our experience, repeating this step for each of the 63 triplets, resulted in the MixITS Interaction Canvas, a design tool to support Interaction Gulf analysis, described in more detail in Section~\ref{sec:canvas}. After a deep reflection phase followed by discussion among the authors, we refined the labels and combined redundant triplets to generate a final set of 36 emerging design pattern triplets reported in Section~\ref{sec:patterns}.

\section{MixIT-Kit Design Toolkit}
\label{sec:themes}
\begin{figure}
    \centering
    \includegraphics[width=1\linewidth]{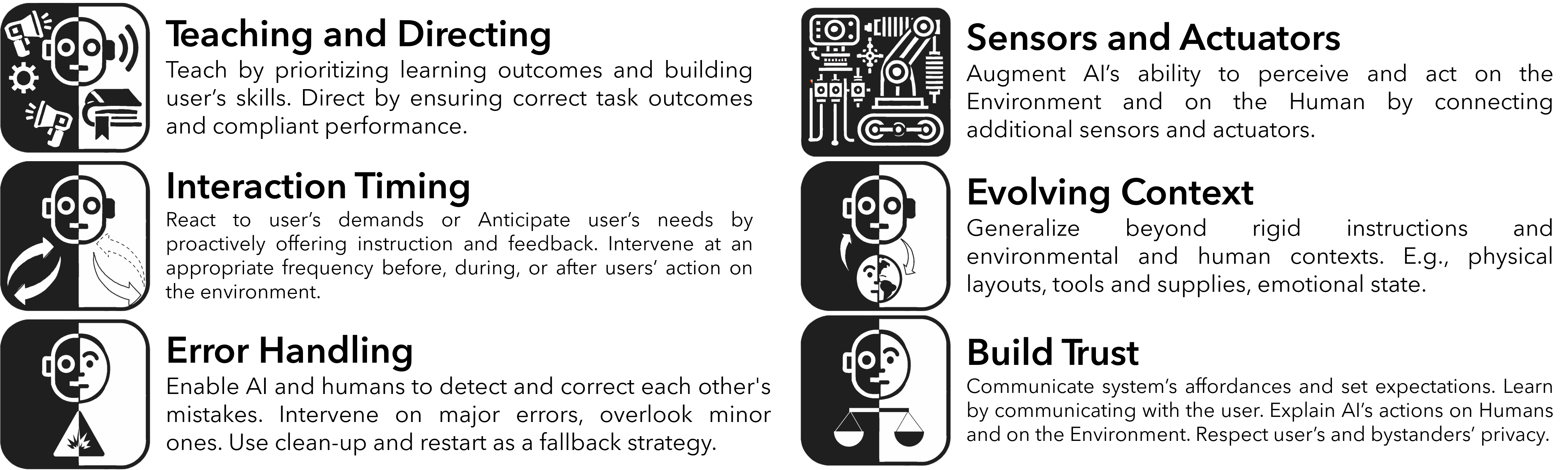}
    \caption{Six \textbf{MixITS Design Considerations} produced by a reflexive thematic analysis of data collected in the design course. Section~\ref{sec:themes} presents the considerations in detail.}
    \label{fig:themes}
\end{figure}

From the analysis of the extensive documentation of the interactive design process of MixITS prototypes during the course, we distilled six design considerations for the design of future MixITS systems. We encourage readers to consider the similarity of their project domain to MixITS (Section~\ref{sec:class_procedure}) to judge their ``proximal similarity''~\cite{campbell1986relabeling} before transferring our findings to their particular cases~\cite{polit2010generalization}. The design considerations are detailed in the following subsections and summarized in Figure~\ref{fig:themes}.

\subsection{Themes and Design Considerations}

The design considerations expand existing human-AI design guidelines by considering adaptive AI, multimodal reasoning, and context-aware interaction. These principles, while drawing from prior AI and MR guidelines, uniquely address real-time physical task support challenges. By emphasizing user state inference, environmental adaptation, and diverse interaction modalities, they bridge digital interfaces and physical tasks. The considerations recognize MixITS systems as operating at the intersection of AI, HCI, and physical performance, requiring nuanced application of cross-domain principles to enhance system efficacy and user experience.

\subsubsection{Teaching and Directing}

We observed that design choices typically aligned with two main goals: teaching skills or directing tasks. Skill-focused MixITS aimed to empower users for independent future ability, implementing pedagogical interventions and assessing learning outcomes. Task-directing MixITS focused on either meeting outcome specifications or guiding users through precise steps. These two approaches differed in their user assumptions and environmental focus.

\textbf{Teaching-focused} MixITS prioritized learning over task completion, sometimes interrupting activities for educational purposes. For example, PianoMix interrupted a user's session to correct technique errors, even when the technique did not adversely affect the musical output. ARCoustic modulated feedback and modeled user emotions to maintain motivation, reflecting research on engagement and learning outcomes ~\cite{harp1997role,erez2002influence}. These designs also allowed users to select instructional preferences and performance metrics self-assessment. Surprisingly, teaching MixITS often lacked safety features and assumed well-controlled environments. They also provided fewer opportunities for users to correct AI mistakes, presuming an ``ideal'' system state and superior AI knowledge.

\textbf{Task-directing} MixITS assumed higher user expertise in real-world settings. These designs allowed users to provide feedback on AI mispredictions (RealityFix and ChefMix), and were more considerate of AI interruption frequency and task disruption. Some systems prioritized task continuity over immediate error correction. User agency was preserved by allowing reasonable improvisation within task steps (CrossReality and ARCoustic). Intervention decisions resembled those proposed by Horvitz principle~\cite{horvitz1999principles}. Safety was a key focus, with designs signaling potential hazards (Cookbot and ChefMix). CrossReality even included the ability to request external medical assistance.

When designing MixITS systems, it is crucial to distinguish between teaching and directing goals, as this choice significantly impacts design decisions. Teaching MixITS should consider prioritizing learning outcomes, ensuring safety even in controlled environments, and incorporating pedagogical interventions~\cite{choi2014effects, laurillard2013teaching} derived from research on Intelligent Tutoring Systems~\cite{burton1982diagnosing, anderson1985intelligent}. In contrast, directing MixITS should consider balancing task continuity with error correction, preserving user agency within task parameters, allowing users to correct AI mistakes, focusing on real-world safety, and document processes for future auditing and improvement, especially in industrial contexts. Both types of systems need to consider AI interruption frequency, user expertise levels, and feedback mechanisms. While incidental learning may occur in task-directing MixITS~\cite{van2006influence,tresselt1960study}, clearly defining the primary goal as either teaching or directing would lead to more optimized and effective designs tailored to their specific purposes.

\subsubsection{Interaction Timing}

MixITS systems were designed with either \textbf{proactive} or \textbf{reactive} interaction styles. Proactive systems offered unsolicited guidance, while reactive ones responded to user requests. This choice was influenced by user expertise, some metric of criticality (e.g., user safety and impact on the task outcome), and perceived user workload. For instance, a system for blind users proactively intervened in dangerous situations (CrossReality), whereas a piano-learning system reactively provided feedback when requested (PianoMix). Students recognized the challenge of balancing proactive interaction frequency with user agency. To address this issue, designs like MusIT, a guitar-teaching prototype, allowed users to adjust the intervention frequency according to their preferences. Students designed interventions for three key phases: before, during, and after tasks. Before-task interventions typically included feature tutorials (ChefMix) and procedure summaries (Cookbot and ARCoustic). During-task interventions ranged from minimal guidance (PianoMix and ARCoustic) to critical mistakes or safety alerts (\textit{Cookbot}). Post-task interactions offered performance reports and outcome assessments. To address interaction frequency concerns, some designs, like MusIT, accumulated recommendations for periodic or end-of-session reports, demonstrating how temporal alignment can optimize user experience.

Designers should consider balancing proactive and reactive interactions and their frequency based on user expertise, task criticality, and workload. Simultaneously, they need to be mindful of how interaction frequency impacts user workload, fatigue, and error rates. A flexible system would allow users to adjust AI intervention frequency and tailor interactions to specific task phases, which can help personalize the experience and retain user agency.

Optimizing interaction frequency can avoid disrupting task flow while maintaining system usefulness. System actions, if aligned with task execution, would minimize disturbance, and implementing feedforward techniques for ``before-task'' interactions would help users anticipate consequences. Interactions that are too infrequent might lower the system's usefulness with suboptimal results and inefficiencies. This resonates with the guideline to ``time services based on context'' by Amershi et al.~\cite{amershi2019guidelines} for non-MixITS systems. Existing guidelines on mixed-initiative systems~\cite{st1997interaction, allen1999mixed}, utility concepts~\cite{horvitz1999principles}, and contextual timing of services offer valuable insights that complement our design considerations. The implications of unsolicited assistance for learning have been explored in the intelligent tutoring systems domain as the ``assistance dilemma''~\cite{koedinger2007exploring}. Frameworks such as XAIR~\cite{xu2023xair} can provide specific guidance on timing AI explainability effectively.

\subsubsection{Error Handling}
Students identified errors in user-environment interactions via a simulated walkthrough with a task flowchart analysis that highlighted steps, outcomes, and required resources. Cognitive walkthroughs and prototype testing revealed potential errors in user-AI interactions and AI's perception of the environment. This process uncovered potential errors in MR interface interactions, AI predictions, and environmental changes. Four themes emerged from students' error-handling strategies.

\paragraph{Prevention}

Prototypes demonstrated strategies involving users in decision-making. MusIT incorporated a feedback loop where the AI prompted the user for confirmation of critical information before performing tasks like optical character recognition (OCR) on user-shared documents. ChefMix had two error prevention features: one allowing users to notify the system about unfulfilled step preconditions before executing an AI-instructed step, and another where the AI preemptively detected unfulfilled preconditions overlooked by users. These approaches showcase collaborative user-AI error mitigation methods.

\paragraph{True Positives}
Teams implemented various strategies for handling AI and user errors. When AI detected its own mistakes, RealityFix acknowledged and apologized, while Cookbot implemented a collaborative problem-solving approach initiated by the user describing the problem. For user errors detected by AI, PianoMix and ARCoustic could interrupt tasks and provide immediate corrections, ensuring proper practice. ChefMix went further by incorporating emotional state management before guiding users through corrections, aiming to reduce frustration and enhance the learning experience. These approaches indicate the goal to increase transparency, build user trust, and improve error handling in AI-user interactions.

\paragraph{False Positives}

We identified strategies for handling AI false positives in error detection. PianoMix enabled users to dismiss incorrectly identified errors, maintaining user control over task disruptions. ARCoustic implemented an interactive approach where users could counter-argue AI-detected mistakes and add an explanation to correct the AI's understanding, offering a way to override the system's false positives and provide data for future refinement of the algorithm. Notably, no features were identified where AI could judge user-detected errors as false positives, suggesting a prioritization of user agency in error handling. These approaches indicate a balance between AI assistance and user control in error management.

\paragraph{Intractable Errors}

Teams devised various strategies to handle errors that went undetected by both users and AI during the task. Cookbot offered a user report feature for post-task issue reporting to developers, which can assist future system updates. RealityScan had an AI feature supporting users in deductive thinking, enhancing error log quality through collaborative error detection. Fail-safe mechanisms were included on Cookbot and ChefMix, allowing users to manually shut down the system, clean the work area, and start again. CrossReality focused on user safety, incorporating shortcuts and safeguards to trigger external support like medical help when necessary. 

\paragraph{Design Considerations}

Designers should consider prioritizing error prevention and comprehensive error handling. This could be done by incorporating proactive strategies, such as feedback loops for user confirmation of critical information and preemptive detection of unfulfilled preconditions for a specific task. Feedforward techniques that allow users to anticipate the consequences of their actions have been shown to significantly decrease errors by enhancing user understanding of the system’s behavior and potential outcomes~\cite{coppers2019fortunettes, muresan2023using, artizzu2024virgilites}. Error handling mechanisms need to address both AI and user mistakes, including AI acknowledgment of its errors, user-initiated problem-solving, AI interruption for immediate user error correction, and related impact on learning and flow. Emotional regulation features can enhance the learning experience by reducing user frustration during corrections. The design should consider user agency in dismissing false positives and the ability to provide feedback to the AI without making it burdensome. Post-task error reporting could help address undetected issues while helping improve the overall system. Depending on the context, fail-safe mechanisms like manual system resets and safety triggers for external support can be integrated. Overall, there is need for designs to balance AI assistance with user control, prioritize transparency to build trust, and ensure adaptability to various error scenarios in physical task contexts. With a human-in-the-loop, designers can enable users to guide and correct AI behavior in realtime, thus enhancing the system's adaptability and reliability over time. Designers can additionally leverage prior research on human errors that provides useful taxonomies and approaches to handle skill-based, knowledge-based, and perceptual errors~\cite{shappell2000human, norman1981categorization}.

\subsubsection{Sensors and Actuators}

Some MixITS prototypes showcased enhanced sensing and actuating capabilities beyond standard MR headsets. ChefMix installed fixed cameras in the environment, providing a third-person view to reduce occlusions and improve physical environment modeling. CrossReality, focusing on assisting blind users in street crossing at a stop sign, developed a system that coordinated with smart city devices like traffic lights and vehicles. Their prototype included wearable devices using haptic stimuli for guidance and an augmented cane with buttons for user input, enabling effective interactions in noisy environments without relying on audio feedback. These enhancements show the potential for MixITS systems to integrate with broader infrastructures and adapt to diverse user needs and environmental challenges.

Future MixITS systems should consider integrating IoT and robotics to enhance environmental interaction capabilities. Augmented tools with embedded sensors could provide precise task guidance, while leveraging existing 3D mapping data could enhance spatial awareness. A wearable ecosystem could provide physiological data for personalized guidance, and networked expertise could offer real-time access to specialized knowledge and potentially reduce AI errors. Bodily control technologies like Electrical Muscle Stimulation (EMS)~\cite{nith2024splitbody, shahus2023skillab} and Galvanic Vestibular Stimulation (GVS)~\cite{sra2017galvr, byrne2016balance} could add to user path guidance, potentially reducing workload and enhancing performance. Designers need to address challenges such as sensor obstruction, particularly as MR headsets block facial expressions, necessitating alternative methods for emotional state inference. Leveraging existing technologies and infrastructure can help overcome current limitations in AI and MR technologies, enabling more sophisticated user-environment interactions in task guidance.

\subsubsection{Evolving Context}

One of the major benefits of MixITS systems over other types of guidance such as video tutorials, beyond realtime feedback, is their potential for adaptability. MixITS systems are expected to generalize beyond rigid instructions and seamlessly adapt to the user's context and ability. 
ChefMix and ARCoustic used human reactions as a shortcut to infer contextual changes, bypassing pure machine perception. They detected important changes through shifts in facial expressions and gaze patterns, prompting proactive system responses. The prototypes also adapted their interaction modes to environmental changes. For instance, ARCoustic typically provided feedback after listening to a user's guitar performance, but in noisy settings, it temporarily switched to demonstrating song passages and allowing for user self-assessment. 

AI can allow systems to learn and refine based on user experience and feedback~\cite{ouyang2022training}. When designing MixITS systems, designers should consider: implementing adaptive learning based on user feedback; utilizing multimodal AI to process various data types for comprehensive reasoning; inferring user workload and interaction capacity through multimodal analysis~\cite{liu2024human}; detecting changes in manipulated objects for contextual adaptation; and leveraging diverse hand grasps to enable a range of interaction possibilities ranging from single-finger gestures to graphic grasp-centered interfaces~\cite{sharma2021solofinger, aponte2024grav}. These considerations allow for more responsive, context-aware systems that can adjust their guidance methods based on user needs and environmental factors.

\subsubsection{Building Trust}
In the context of physical task guidance, research indicates the importance of trust for effective human-AI interaction~\cite{haesler2018seeing, eyck2006effect}. Students addressed this by incorporating trust-building features into their prototypes. Cookbot, ARCoustic, and MusIT allowed users to report AI mistakes, enabling system adaptation and fostering a sense of user influence. ARCoustic and MusIT explained errors and recommendations to the users, while Cookbot and ChefMix used MR visuals to show tracked environmental elements, enhancing transparency. These approaches, aligned with frameworks like XAIR for explainable AI in AR \cite{xu2023xair}, aimed to establish two-way communication between users and the system, to foster long-term trust development. Explainability is a key factor in trust-building, that aligns with Amershi et al.'s guideline to clarify system actions \cite{amershi2019guidelines}. Cookbot and ARCoustic visualized task progress and allowed users to fast-forward or return to a specific step of the task procedure, enhancing user control. Addressing privacy concerns, students recognized the sensitive nature of multimodal data collection in MixITS systems. PianoMix implemented informed consent procedures, while other projects proposed visual indicators for active data collection and measures to protect bystander privacy, demonstrating awareness of ethical considerations in AI-driven task guidance.


While not unique to MixITS systems, designers should focus on building trust through transparent decision-making, error reporting, and system adaptation. Explainable AI using visual cues to demonstrate awareness and reasoning, should be considered. Enhancing user control with task navigation and progress visualization can help with user agency and associated trust. Prioritizing privacy through clear consent procedures and data collection indicators can help users feel comfortable using a MixITS system. Clearly communicating system capabilities and limitations can help set accurate user expectations, particularly as questions about artificial general intelligence arise~\cite{terry2023ai}. This consideration resonates with the guideline to ``Make clear what the system can do'' from Amershi et al.~\cite{amershi2019guidelines}. Designing adaptive systems that adjust guidance methods based on user feedback and environmental factors can help ground the systems into the user's immediate context. These considerations can create trustworthy, transparent, and effective MixITS systems that balance advanced capabilities with user-centric design.

\subsection{Gulfs of Execution and Evaluation in Intelligent Mixed Reality}\label{sec:gulfs}

\begin{figure}
    \centering
    \includegraphics[width=.8\linewidth]{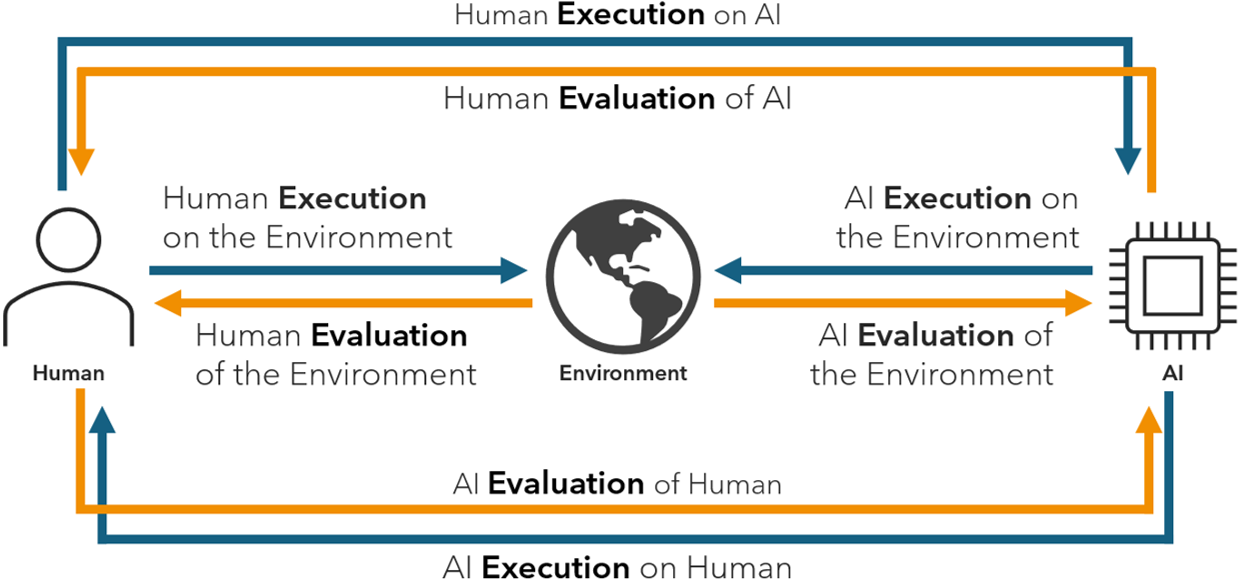}
    \caption{When applying Don Norman's Gulfs of Execution and Evaluation~\cite{norman1986user, norman2013design} to MixITS systems, we identify eight Gulfs that occur during interactions between the Human user, AI-MR system, and the Real Environment.}
    \label{fig:gulfs}
\end{figure}

Donald Norman’s Gulfs of Evaluation and Execution~\cite{hutchins1985direct, norman1986user, norman2013design} have been instrumental for HCI researchers and practitioners in understanding the challenges users encounter when interacting with systems~\cite{vermeulen2013crossing, hornbaek2017interaction, muresan2023using}. The gulf of execution refers to the gap between a user's intended action and the options a system provides to perform that action. A small gulf indicates that the system's controls align well with the user's intentions, making it easy to use. A large gulf of execution means the system's interface poorly matches the user's goals, making it difficult to accomplish tasks.

The Gulf of Evaluation is the gap between how a system presents its state and how a user interprets it. A small gulf of evaluation means users easily understand the system's feedback and current state. A large gulf of evaluation indicates users struggle to accurately interpret the system's state, leading to confusion or errors.

We apply Norman's gulfs to identify and analyze interactions in a MixITS system, which includes the new element of the physical environment and changes the system from the traditional deterministic system to an AI-based probabilistic one.
In a MixITS system, users need to convey their intentions to both the AI (system) and the physical environment simultaneously. This dual interaction can create two distinct gulfs: (1) Gulf of Human Execution on AI (\textbf{H-Ex-AI}) occurs when users struggle to communicate their intentions to the AI system, and (2) Gulf of Human Execution on Environment (\textbf{H-Ex-E}) which arises when users face challenges in physically executing their intentions in the real world. For example, in an AI-assisted rock climbing system, a climber might intend to reach a misidentified hold on a path suggested by the AI. They need to indicate this mistake to the AI (H-Ex-AI) to allow it to update its suggested path, perhaps through gaze or gesture, while also physically moving their body to grasp a different hold (H-Ex-E) that is more within reach.

After taking action, users must interpret feedback from both the system and the physical environment. This dual interpretation can lead to two distinct gulfs: (1) Gulf of Human Evaluation of AI (\textbf{H-Ev-AI)} which occurs when users struggle to understand or interpret the feedback provided by the AI system; and (2) Gulf of Human Evaluation of Environment (\textbf{H-Ev-E)} which arises when users face challenges in perceiving or interpreting the results of their actions in the physical world. Continuing the rock climbing example, after reaching for a hold, the climber needs to process the AI's feedback about their technique or next move (H-Ev-AI), perhaps displayed through MR visuals or audio feedback. Simultaneously, they must evaluate their physical position and stability on the wall (H-Ev-E).

We can consider MixITS systems as mixed-initiative systems~\cite{horvitz1999principles} where the AI backend can behave as an agent and initiate interactions with the user and the physical environment. This mixed-initiative perspective leads to four additional Gulfs: Gulf of AI Execution on Human (\textbf{AI-Ex-H}), Gulf of AI Execution on the Environment (\textbf{AI-Ex-En}), Gulf of AI Evaluation of Human (\textbf{AI-Ev-H}), and Gulf of AI Evaluation of the Environment (\textbf{AI-Ev-E}). Figure~\ref{fig:gulfs} shows all the eight Gulfs of a MixITS application.

We treat the physical environment as a passive entity affected by user and AI actions, rather than as an active participant. This simplification keeps our design toolkit focused and practical, avoiding the complexities of fully modeling environmental factors.

\subsection{Design Patterns}
\label{sec:patterns}
\renewcommand{\arraystretch}{2.25}
\begin{table*}
\centering
\caption{
The initial 36 \textbf{PTG Design Patterns} in the Problem-Solution format tuple and grouped by Interaction Gulfs.}

\resizebox{.925\textwidth}{!}{%
\begin{tabular}[t]{p{100px} l l l}

\textbf{Gulf} & \textbf{ID} & \textbf{Problem} & \textbf{Example Solutions} \\

\Xhline{1.5pt}
\multirow[t]{20}{*}[-2em]{
    \shortstack[l]{
        \textbf{H-Ex-E} \\
        \textit{Human Execution} \\
        \textit{on the Environment}
    }
}
\\[-2em]

&
1.
&
Environment Understanding
&
\multirow[t]{2}{*}[-.9em]{
    \shortstack[l]{
        Provide additional information about a real environment element.\\
        \textit{E.g., descriptive labels registered to unfamiliar components of a PC.}
    }
}
\\ [5px] 

&
2.
&
Risk Awareness
&
\multirow[t]{2}{*}[-.9em]{
    \shortstack[l]{
        Highlight potentially dangerous real environment elements before the user executes the step.\\
        \textit{E.g., MR arrows and audio warning about a hot pan nearby.}
    }
}
\\ [5px] 

&
3.
&
Step Preconditions
&
\multirow[t]{2}{*}[-.9em]{
    \shortstack[l]{
        Notify the user about an unfulfilled precondition of the next step.\\
        \textit{E.g., A spoken instruction to the user to tune their guitar before starting practice.}
    }
}
\\ [5px] 

&
4.
&
Procedural Knowledge
&
\multirow[t]{2}{*}[-.9em]{
    \shortstack[l]{
        Provide instructions with written text and speech.\\
        \textit{E.g., Speech synthesis and MR text/visuals instructing the user on how to operate a rice cooker.}
    }
}
\\ [5px] 

&
5.
&
Spatial Knowledge
&
\multirow[t]{2}{*}[-.9em]{
    \shortstack[l]{
        Visually demonstrate the trajectory of an action or the direction of the relationship between two elements.\\
        \textit{E.g., Animated MR arrow moving between sockets to guide PC cable re-connection.}
    }
}
\\ [5px] 

&
6.
&
Body Movement
&
\multirow[t]{2}{*}[-.9em]{
    \shortstack[l]{
        Visually demonstrate body movement or pose with an animated 3D body, video, or image.\\
        \textit{E.g., Animated MR hand to demonstrate how to play a lead passage on the piano.}
    }
}
\\ [5px] 
 
&
7.
&
High User Workload
&
\multirow[t]{2}{*}[-.9em]{
    \shortstack[l]{
        Provide virtual functionalities that aid in the task execution.\\
        \textit{E.g., Timer to keep track of cooking time.}
    }
}
\\ [5px] 

&
8.
&
Virtual Environment Clutter
&
\multirow[t]{2}{*}[-.9em]{
    \shortstack[l]{
        Hide virtual overlay.\\
        \textit{E.g., A voice command to hide or dim MR labels on objects in a repair shop.}
    }
}
\\ [5px] 

&
9.
&
Real Environment Clutter
&
\multirow[t]{2}{*}[-.9em]{
    \shortstack[l]{
        Indicate real environment elements that can be removed from the workspace.\\
        \textit{E.g., Register MR labels to indicate a utensil is not needed for a recipe and can be stored.}
    }
}
\\ [5px] 

&
10.
&
Lack of Focus
&
\multirow[t]{2}{*}[-.9em]{
    \shortstack[l]{
        Highlight a single real environment element.\\
        \textit{E.g., Complement instructions by pointing an MR arrow to a PC component that should be connected.}
    }
}   
\\ [20px]

\Xhline{1.5pt}
\multirow[t]{20}{*}[-2em]{
    \shortstack[l]{
        \textbf{H-Ev-E} \\
        \textit{Human Evaluation} \\
        \textit{on the Environment}
    }
}
\\[-2em]

&
11.
&
Expected Task Result
&
\multirow[t]{2}{*}[-.9em]{
    \shortstack[l]{
        Describe the expected outcome of an entire task.\\
        \textit{E.g., Image of the dish the user wants to prepare.}
    }
}
\\ [5px] 

&
12.
&
Task Result
&
\multirow[t]{2}{*}[-.9em]{
    \shortstack[l]{
        Provide a comprehensive report of the final product of the task, explaining inconsistencies and suggesting corrective actions.\\
        \textit{E.g. Superimposed MR model of a correctly assembled PC over the user's assembly to visually highlight discrepancies.}
    }
}
\\ [5px] 

&
13.
&
Task Performance
&
\multirow[t]{2}{*}[-.9em]{
    \shortstack[l]{
        Provide a comprehensive report of the user performance throughout the task, explaining mistakes and suggesting technique improvements.\\
        \textit{Line plot on an MR panel showing user tempo accuracy, highlighting mistakes with audio samples, and suggesting tailored training for those passages.}
    }
}       
\\ [5px] 

&
14.
&
Task Progress
&
\multirow[t]{2}{*}[-.9em]{
    \shortstack[l]{
        Inform task completion progress.\\
        \textit{E.g. MR panel with a checklist of steps.}
    }
}
\\ [5px] 
 
&
15.
&
Expected Step Results
&
\multirow[t]{2}{*}[-.9em]{
    \shortstack[l]{
        Describe the expected outcome of a step of the task.\\
        \textit{E.g., Image of herbs diced or chiffonade as required by the recipe.}
    }
}
\\ [5px] 
 
&
16.
&
Step Results
&
\multirow[t]{2}{*}[-.9em]{
    \shortstack[l]{
        Inform the user of an inconsistency in the step outcome and recommend correction based on the impact.\\
        \textit{E.g., Interrupt (if allowed) the user during incorrect chord play and show an image of the correct music sheet.}
    }
}
\\ [5px] 
 
&
17.
&
Tool Operation
&
\multirow[t]{2}{*}[-.9em]{
    \shortstack[l]{
        Demonstrate the correct operation of the physical tool and contextualize mistakes by overlaying content on the tool.\\ \textit{E.g., Overlay MR numbered labels on the guitar neck to display the correct chord shape.}
    }
}
\\ [5px] 
 
&
18.
&
Causal Chain
&
\multirow[t]{2}{*}[-.9em]{
    \shortstack[l]{
        Review previous step records with the user to identify undetected circumstances that caused inconsistencies in outcomes.\\
        \textit{E.g., Engage in a conversation with the user to reason about pieces of evidence after analyzing a crime scene.}
    }
}
\\ [5px] 
 
&
19.
&
User Frustration
&
\multirow[t]{2}{*}[-.9em]{
    \shortstack[l]{
        Reassure the user, offer instructions to correct performance or outcome issues, and alternative courses of action.\\ \textit{E.g., Synthesize reassuring speech and corrective measures in a calming tone after the user drops a bowl of ingredients.}
    }
}    
\\ [5px] 
 
&
20.
&
Spatial Knowledge
&
\multirow[t]{2}{*}[-.9em]{
    \shortstack[l]{
        Visually demonstrate the trajectory of a previously executed step or the relationship between two elements involved in a previous step.\\
        \textit{E.g., Demonstrate the trajectory a suspect may have followed to enter a crime scene based on the evidence.}
    }
}
\\ [20px]

\Xhline{1.5pt}
\multirow[t]{20}{*}[-2em]{
    \shortstack[l]{
        \textbf{H-Ex-AI} \\
        \textit{Human Execution} \\
        \textit{on AI}
    }
}
\\[-2em]

&
21.
&
Dismissal
&
\multirow[t]{2}{*}[-.9em]{
    \shortstack[l]{
        Allow users to dismiss AI system functionalities easily.\\
        \textit{E.g., ``Thanks, I can take it from here.'' -- User.}
    }
}
\\ [5px] 
 
&
22.
&
Activation
&
\multirow[t]{2}{*}[-.9em]{
    \shortstack[l]{
        Allow users to request AI system functionalities easily.\\
        \textit{E.g., ``CookGPT (activation keyword), how much olive oil should I use?'' -- User.}
    }
}
\\ [5px] 
 
&
23.
&
Scoped Activation
&
\multirow[t]{2}{*}[-.9em]{
    \shortstack[l]{
        Allow users to easily request AI system functionalities directly related to a virtual or real environment element.\\
        \textit{E.g., MR buttons, linked to music sheet sections, offering guidance upon tapping.}
    }
}
\\ [5px] 
 
&
24.
&
Task Obstruction
&
\multirow[t]{2}{*}[-.9em]{
    \shortstack[l]{
        Support interaction modalities that do not obstruct user interactions with the real environment.\\
        \textit{E.g., speech, eye gaze, adaptive UI, tangible inputs.}
    }
}
\\ [20px]

\Xhline{1.5pt}
\multirow[t]{20}{*}[-2em]{
    \shortstack[l]{
        \textbf{H-Ev-AI} \\
        \textit{Human Evaluation} \\
        \textit{of AI}
    }
}
\\[-2em]
 
&
25.
&
Tracking in Progress
&
\multirow[t]{2}{*}[-.9em]{
    \shortstack[l]{
        Inform the user about the parts of the environment being tracked and the expected end time during tracking.\\
        \textit{E.g., MR frame that moves along the environment as the system is scanning and reconstructing specific parts of the environment.}
    }
}
\\ [5px] 
 
&
26.
&
Chain-of-Thought
&
\multirow[t]{2}{*}[-.9em]{
    \shortstack[l]{
        Allow user-system dialogue by explaining the system's decision-making process and enabling user inquiries or counterarguments.\\
        \textit{E.g., Engage in a conversation with the user to explain the reasons why the system is suggesting an alternative theory to explain a crime.}
    }
}
\\ [5px] 
 
&
27.
&
System Focus
&
\multirow[t]{2}{*}[-.9em]{
    \shortstack[l]{
        Inform the user of the system's focus shift to a specific area within the physical environment.\\
        \textit{E.g., An MR icon around the user's hands, while they are mixing ingredients to communicate to the user the system, is aware of the action.}
    }
} \\ [5px] 
 
&
28.
&
Privacy
&
\multirow[t]{2}{*}[-.9em]{
    \shortstack[l]{
        Request explicit user authorization for the acquisition of personally identifiable information.\\
        \textit{E.g., Clearly display the terms of use and obtain explicit user consent to collect data about the user's home.}
    }
}
\\ [20px]

\Xhline{1.5pt}
\multirow[t]{20}{*}[-2em]{
    \shortstack[l]{
        \textbf{AI-Ex-E}\\
        \textit{AI Execution} \\
        \textit{on the Environment}
    }
}
\\[-2em]

&
29.
&
Act on the Environment
&
\multirow[t]{2}{*}[-.9em]{
    \shortstack[l]{
        Coordinate with IoT devices to change the environment towards the user goal.\\
        \textit{E.g., The system requests a smart traffic light, after considering everyone's safety, to turn red so the user can cross the street.}
    }
}
\\ [27.5px]

\Xhline{1.5pt}
\multirow[t]{20}{*}[-2em]{
    \shortstack[l]{
        \textbf{AI-Ev-E} \\
        \textit{AI Evaluation} \\
        \textit{of the Environment}
    }
}
\\[-2em]

&
30.
&
Limited Spatial Awareness
&
\multirow[t]{2}{*}[-.9em]{
    \shortstack[l]{
        Environment modeling from spatial data captured through multi-view RGB, depth-cameras, LiDAR, or millimeter wave radars.\\
        \textit{E.g., A 360-degree camera attached to the MR HMD.}
    }
}
\\ [5px] 
 
&
31.
&
Step Results
&
\multirow[t]{2}{*}[-.9em]{
    \shortstack[l]{
        User feedback to improve the system's assessment of a specific task step outcome locally in the session.\\
        \textit{E.g., The AI misclassifies a chord as a mistake due to background noise, but after the user corrects the system, it recalibrates the microphone.}
    }
}
\\ [5px] 
 
&
32.
&
Procedural Knowledge
&
\multirow[t]{2}{*}[-.9em]{
    \shortstack[l]{
        Allow users to analyze outcomes of past instructions to correct the system's procedural knowledge base.\\
        \textit{E.g., The user informs the system about an incorrect ingredient in the recipe, prompting the system to update the recipe accordingly.}
    }
}
\\ [5px] 
 
&
33.
&
Environment Model
&
\multirow[t]{2}{*}[-.9em]{
    \shortstack[l]{
        Situated user feedback to correct mistakes in the system's environment model.\\
        \textit{E.g., hand gestures to annotate artifacts of misclassification or 3D reconstruction in the system's environment model.}
    }
}
\\ [5px] 
 
&
34.
&
Task Progress
&
\multirow[t]{2}{*}[-.9em]{
    \shortstack[l]{
        User feedback to update task progress in case of undetected step completion.\\
        \textit{E.g., The user informs the system via speech that they have already mixed the necessary ingredients, which was not previously detected by the system.}
    }
}
\\ [20px]

\Xhline{1.5pt}
\multirow[t]{20}{*}[-2em]{
    \shortstack[l]{
        \textbf{AI-Ex-H}\\
        \textit{AI Execution} \\
        \textit{on Human}
    }
}
\\[-2em]

&
35.
&
Instructional Language
&
\multirow[t]{2}{*}[-.9em]{
    \shortstack[l]{
        Rephrase instructions to facilitate understanding when users demonstrate doubt after receiving guidance.\\
        \textit{E.g., In a PC repair scenario, replace a technical term (CPU Cooler) with lay terminology (fan attached to the metallic grid).}
    }
}
\\ [27.5px]

\Xhline{1.5pt}
\multirow[t]{20}{*}[-2em]{
    \shortstack[l]{
        \textbf{AI-Ev-H}\\
        \textit{AI Evaluation} \\
        \textit{of Human}
    }
}
\\[-2em]

&
36.
&
Goal
&
\multirow[t]{2}{*}[-.9em]{
    \shortstack[l]{
        Infer user goals based on explicit inquiries or implicit analysis of their actions in the real environment.\\
        \textit{E.g., Initially, the system assumes the user is playing song A, but upon hearing the first notes, it identifies the song as B and updates the instructions.}
    }
}
\\ [27.5px]

\bottomrule
\bottomrule

\end{tabular}%
} 
\label{tab:full}
\end{table*}

Novice MixITS designers may struggle to apply general AI and MR guidelines to specific projects. Without a shared vocabulary between designers and AI developers, there is a risk of misinterpretations, redundant problem-solving, and inconsistent designs, impacting user memorability and experience ~\cite{winters2009dealing}. We propose design patterns to bridge this gap, offering concrete solutions as examples that can help mitigate the ``cold-start" problem~\cite{winters2009dealing} and facilitate designers and researchers to build upon and expand our set.

We chose not to report pattern frequency to avoid misleading readers to think some patterns are more relevant than others~\cite{maxwell2010using}. Our focus is on establishing a taxonomy of key challenges in MixITS design and suggesting examples of AI and MR solutions. We encourage the reader to determine which patterns are transferable to their needs~\cite{campbell1986relabeling, polit2010generalization}. The 36 patterns, grouped into eight interaction gulfs presented in Table~\ref{tab:full}, offer a foundation for future refinement as the MixITS community grows.

\subsection{Interaction Canvas}
\label{sec:canvas}

\begin{figure}
    \centering
    \includegraphics[width=.8\linewidth]{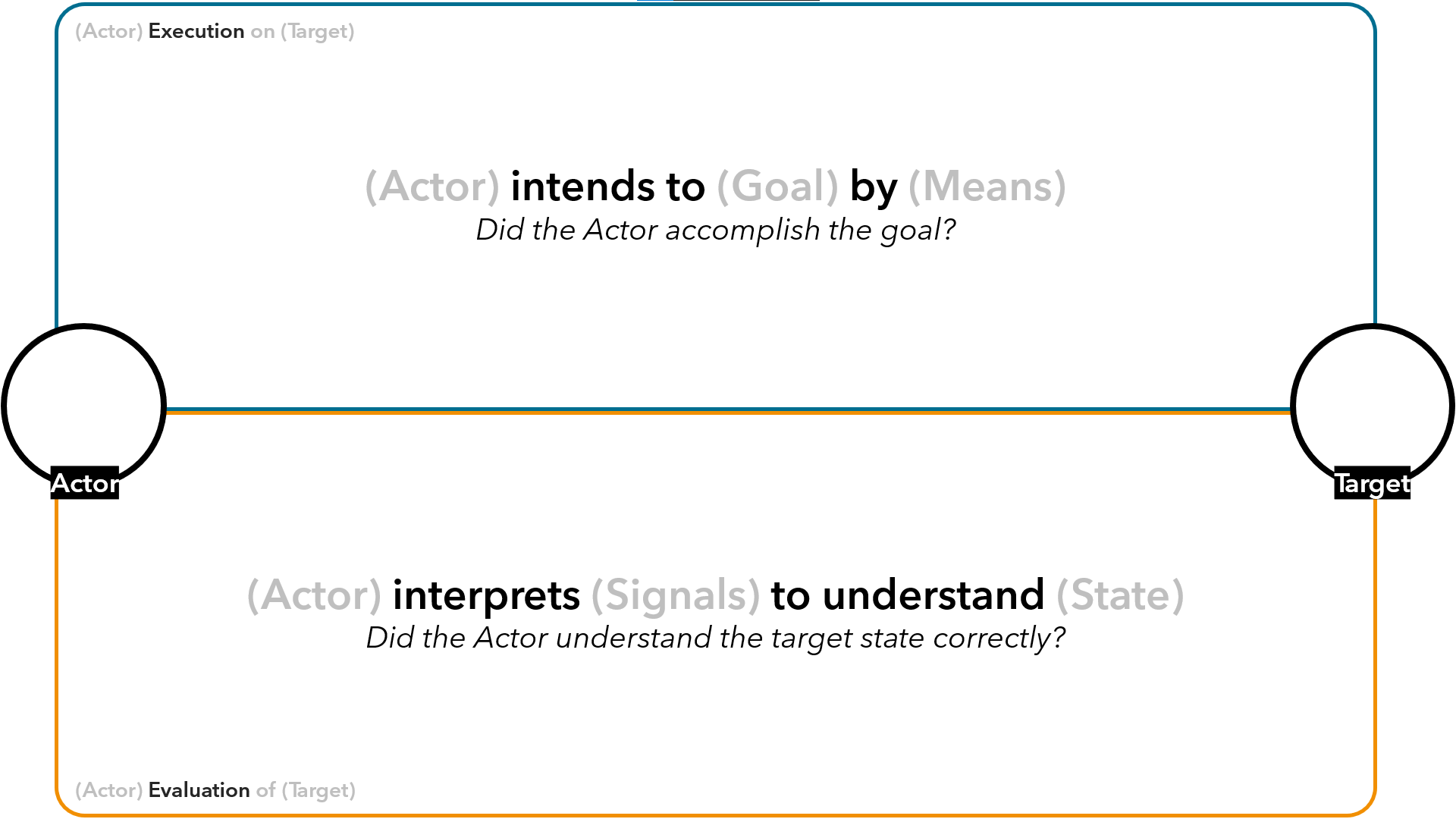}
    \caption{The \textbf{MixITS Interaction Canvas} can help designers identify interaction problems leading to the Gulfs of Execution and Evaluation between the AI system, Human users, and the Environment. Designers can use this visual tool by filling out the blanks, in gray text for both the questions and the gulfs.}
    \label{fig:canvas}
\end{figure}

In the design pattern elicitation process (Section~\ref{subsec:designpattern}), we classified 63 interaction problems into one of eight interaction gulfs (Section~\ref{sec:gulfs}). This exercise led to the creation of the Interaction Canvas (Figure~\ref{fig:canvas}), a visual tool to streamline the analysis of Gulfs of Execution and Evaluation in MixITS systems. We present this Canvas as a tool to help designers visualize their thought processes and communicate effectively when analyzing interaction gulfs in MixITS systems. To use the Canvas, designers fill in the blank spaces at the edges of the canvas and answer: (1) who is the actor initiating the interaction, the AI-MR system or the human? (2) what is the target of the interaction, the AI-MR system, the human or  the environment? Once the actor and target of the interaction are defined, the designer should focus on the execution cycle at the upper half of the Canvas and reflect on: (3) what is the actor's goal in the target? (4) what are the means provided to the actor by the target? (5) did the actor accomplish the goal? Next, designers can focus on the bottom half of the Interaction Canvas and answer: (6) what are the feedback signals emitted by the target? (7) How did the actor interpret those signals? (8) Did the actor understand the target state correctly?

\section{Evaluation}

\begin{figure*}
    \centering
    \includegraphics[width=1\linewidth]{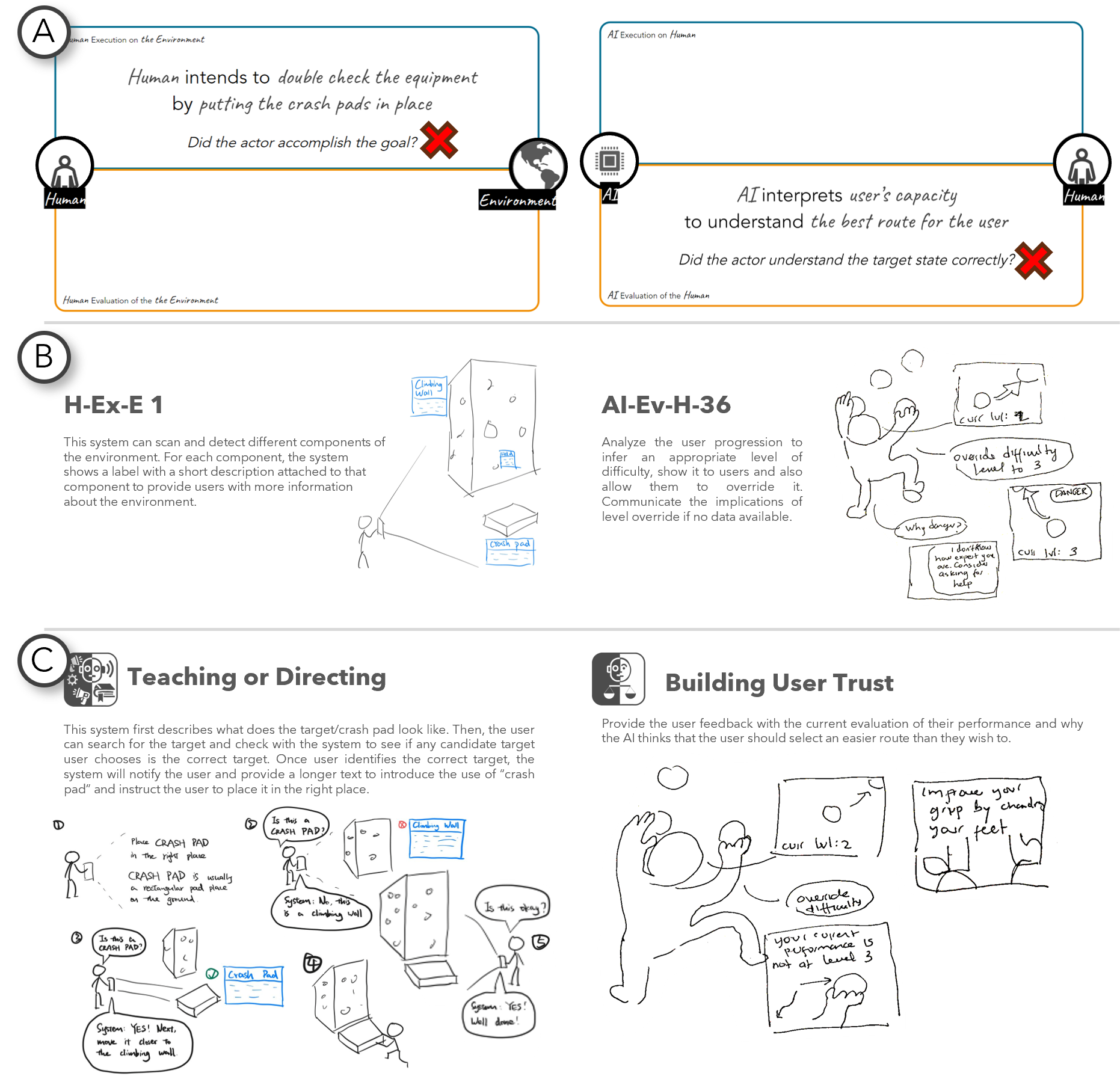}
    \caption{\textbf{(a)} In \textbf{Task 1}, participants used Interaction Canvases to identify a Gulf on fictional-user issues. \textbf{(b)} They then referred to Table~\ref{tab:full} to find a problem that most closely matched their scenario and identified the applicable Design Pattern. \textbf{(c)} In \textbf{Task 3}, they revisited their initial solution through the lens of a Design Consideration. P6 used descriptive AR labels (H-Ex-E-1) with didactic instructions on the importance of crashpads and their use, considering ``Teaching and Directing''. After reflecting on ``Build Trust'', P10 added an AI explanation to the goal inferred by AI (AI-Ev-H-36).}
    \label{fig:eval_samples}
\end{figure*}

To evaluate MixITS-Kit, which comprises the six design considerations, 36 design patterns, and the interaction canvas, we conducted an asynchronous take-home study~\cite{ledo2018evaluation} with eight participants. They were required to complete three tasks using the different components of MixITS-Kit to improve a fictional MixITS app for novice indoor rock climbers (ClimbAR). We gathered data from their work, including artifacts, feedback, and survey responses. The study was approved by our local IRB and participants provided informed consent before proceeding with the study tasks. Participants were compensated for their time.

\subsection{Participants}

We recruited eight participants (4F, 3M, 1NB; age range 18-33), all students from different departments of our home university who had prior experience with MR and AI. None of the participants were part of the human-AI interaction design course used for our initial data collection. We consider this participants are representative the target audience of MixITS-Kit, early career engineers and designers who have some technical experinece but are new to the MixITS domain.

One was a master's student, two were PhD students, and five were undergraduate students. Participants self-reported their expertise levels in designing AI, MR, and MixITS systems using a Likert scale (1 = ``Never designed such a system'', 5 = ``Expert''). The mean self-reported expertise levels were as follows: 2.4 in AI (SD = 1.1), 2.7 in MR (SD = 1.2), and 2.0 in MixITS (SD = 1.0). We asked participants about their awareness of existing design toolkits for AI, MR, and MixITS. For AI, the most frequently mentioned toolkits were PyTorch (mentioned four times), TensorFlow, scikit-learn, and the Gemini API (each mentioned twice). For MR, the most commonly mentioned toolkits were MRTK (four times) and Unity AR Foundation (5 times). When asked about design toolkits for MixITS, participants either did not mention any (five times) or mentioned technical MR toolkits (three times).

\subsection{Apparatus} The study was conducted remotely and asynchronously. Participants completed the tasks using a popular online slide authoring tool. We shared a digital version of our MixITS-Kit with participants using the same online platform. The Interaction Canvas was shared as a slide embedded in the main study slide deck containing Figure~\ref{fig:canvas}, the design patterns were shared as a PDF version of Table~\ref{tab:full}, and the design considerations were shared as slides with visuals and summaries from Section~\ref{sec:themes}. We collected feedback and surveys using forms in the same online platform.

\subsection{Procedure}
Our procedure combines a walkthrough demonstration with a take-home study, similar to previous toolkit evaluations~\cite{ledo2018evaluation}. We instructed participants to to work directly on the slides so all the changes were tracked. Each of the eight participants were randomly assigned to a unique condition starting at one of the eight MixITS gulfs and later encountering two other conditions varying gulf direction and actors least one change in the gulf direction and actor. This balancing strategy ensured every participant encountered every actor and every gulf direction.

Before participants proceeded with the actual study tasks, they were presented with introductory slides. These included one slide about Don Norman's principles of design~\cite{norman2013design}, one slide with the definition of MixITS, and one slide describing the tools in our MixITS-Kit. Following this introduction, additional slides introduced the fictional setting of the study. This warm-up phase concluded with thirteen slides providing a \textbf{walkthrough demonstration}~\cite{ledo2018evaluation} of how to accomplish a task similar to the upcoming task using MixITS-Kit. Each participant received a unique warm-up example different from the subsequent experimental tasks.

In the first task (\textbf{Task 1}), participants received a fictional error reported by the user. We created errors contextualized in the experimental scenario and based on one of the eight MixITS interaction gulfs according to the condition (Table~\ref{tab:full}). They described the user's issue by filling out the canvas (Fig.~\ref{fig:canvas}), selected and justified a relevant interaction problem from Table~\ref{tab:full}, and proposed a solution using the associated design pattern. They detailed the solution by including text descriptions and a sketch image. This task assessed whether participants could effectively navigate design pattern catalog to identify a similar problem, use the canvas to analyze that problem, and apply a suitable design pattern to solve the reported issue.

In the next task (\textbf{Task 2}), participants reviewed Task 1 solutions from other participants, ensuring they faced a scenario not previously encountered in either Task 1 or the warm-up phase. Given only the solution text description and sketch image, they had to identify which design pattern from Table~\ref{tab:full} their peer applied, which then would help identify the associated gulf and problem. This exercise aimed to assess whether participants could recognize patterns in their peer's work, indicating the potential of MixITS-Kit as a shared language for MixITS designers and developers.

In the final task (\textbf{Task 3}), participants revised their Task 1 solution using a design consideration (Section~\ref{sec:themes}) most relevant to their original scenario. All considerations were used at least once, with ``Interaction Timing'' and ``Error Handling'' used twice due to their complexity. This task aimed to explore how the design considerations can influence and creatively expand MixITS solutions. Participants completed a survey after this task.

\subsection{Data Collection}

In the productive tasks (Tasks 1 and 3), participants were asked to rate their level of agreement with various statements using a 5-point Likert scale. The statements focused on Learnability (\textit{``It was easy to learn how to use the toolkit.''}), the toolkit's effectiveness as a shared vocabulary (\textit{``I can easily communicate using the vocabulary of the toolkit.''}, \textit{``I can easily understand the vocabulary of the toolkit.''}), and the research goals of the toolkit as proposed by Ledo et al.~\cite{ledo2018evaluation} (\textit{``It would take me longer to solve the task without the toolkit,'' ``The toolkit helped me to identify paths of least resistance in the design process,'' ``The toolkit will help to empower future designers,'' ``The toolkit integrates well with current practices in design,'' and ``The toolkit helped me to create a novel design.''}). We also asked participants whether they found the toolkit too abstract and high-level, too low-level, or at an appropriate level of abstraction for productive tasks 1 and 3. We measured task completion time by summing the time intervals participants spent on tasks 1 and 3, as recorded in the slide change history. Task 2 evaluated participants' ability to correctly identify design patterns used by other participants. Therefore, we recorded only the names of recognized patterns and whether they matched the originally intended ones.

\subsection{Data Analysis}
We analyzed the medians and median absolute deviations of the questionnaire responses and task completion times. Additionally, we examined the ratios of responses from the questionnaires, toolkit abstraction level feedback, and the correct recognition ratio from Task 2. One of the authors analyzed the solution descriptions and visuals from tasks 1 and 3 using the same pattern mining method employed in Section~\ref{sec:patterns}.

\subsection{Results}

\begin{figure}
    \centering
    \includegraphics[width=.75\linewidth]{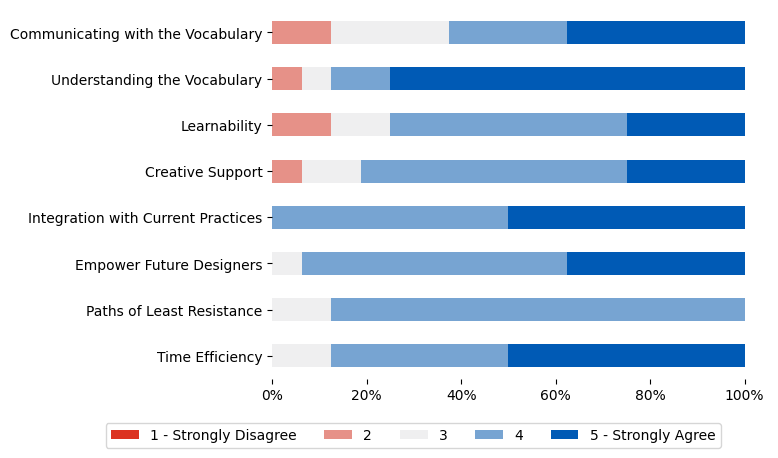}
    \caption{Participant agreement levels (5-point Likert scale) with statements about the Design Toolkit.}
    \label{fig:likert}
\end{figure}

The median task completion times in minutes were: Task 1 (M=27.0, MAD=9.0), Task 2 (M=1.0, MAD=0.5), and Task 3 (M=8.5, MAD=3.5). Participants generally responded positively to MixITS-Kit, as shown in Figure~\ref{fig:likert}. Participants recognized the potential of MixITS-Kit to serve as a shared vocabulary among designers. However, they found it harder to communicate using the vocabulary (M=4, MAD=1) than to understand it (M=5.0, MAD=0.0). Participants considered it easy to learn MixITS-Kit (M=4.0, MAD=0.5). Regarding the level of agreement with sentences related to Ledo et al.'s~\cite{ledo2018evaluation} toolkit research goals, participants rated creative support (M=4.0, MAD=0.0), integration with current design practices (M=4.5, MAD=0.5), empowering designers (M=4.0, MAD=0.0), supporting paths of least resistance (M=4.0, MAD=0.0), time-efficiency (M=4.5, MAD=0.5). Regarding the level of abstraction, one of the participants (P9) considered the  toolkit too abstract for the task (12.5\%), in contrast to the other seven, who considered the components to be at appropriate levels of abstraction (87.5\%).

In Task 1, two of the eight participants applied design patterns from gulfs different than anticipated and provided justifications that did not align with the defined gulf concepts, indicating a mistake.

In Task 2, all participants correctly identified the actor and target in their peers' solutions. Four participants successfully recognized the correct design pattern, while five identified the correct gulf. One participant misidentified the pattern, and three incorrectly identified the gulf. Overall, half of the participants accurately identified the exact design pattern used in their peer's solution.

Participants identified the most challenging aspects of learning the canvas and design patterns for Task 1, including\textit{``It was harder to identify the gulf.''} - P9 and \textit{``Understand the mapping between the gulf concepts and the terms in the toolkit ''} - P10. When asked for open-ended feedback on the canvas and design patterns in Task 1, participants said, \textit{``I thought the toolkit was very useful to standardize problems in AI-MR apps, and the design patterns seem pretty comprehensive to me.''} - P1, \textit{``adding an `Other' section would allow developers to potentially communicate about rarer cases''} - P4, \textit{``An interface that integrates both parts would make it easier to cross-reference.''} - P6, \textit{``I used the problem list to identify a problem that matched that one in the case, and then that helped me understand better how to fill each gulf''} - P10.

When asked for open-ended feedback on the design consideration in Task 3, participants said, \textit{``I think Task 2 and 3 were a lot more straightforward than the first task! After gaining familiarity with the design recommendation, I also found it easier to proceed.''} - P8, \textit{``Maybe it's just me being not creative enough, but it's still a little bit difficult to come up with a very novel design using the guidance in my scenario''} - P7 (``Sensors and Actuators'' and Human Execution on the Environment), \textit{``I used the design recommendation to find a pattern that aligned with the goal stated on it.''} - P6, \textit{``I thought the design recommendations were great guidelines to help find solutions to the proposed problems. It was especially interesting that they addressed both the AI and XR aspects of design, since most toolkits I've seen/used only work with one of those aspects.''} - P1.

\section{Discussion}
MixITS-Kit consists of three interconnected components: an Interaction Canvas for analyzing execution and evaluation gulfs, a set of six user-centered design considerations, and a set of 36 low-level design patterns with example solutions. This structure can support designers throughout the development process, from conceptual analysis to practical implementation. Here we discuss our results, the synergy between the toolkit components, and their overall implications for the landscape of HAI.

\subsection{Positive Impacts of MixITS-Kit}

One of the main findings from our evaluation was the high level of performance demonstrated by participants. Six out of eight participants successfully completed Task 1. In Task 2, all participants correctly identified the actors and targets, while half accurately recognized the design pattern used by their peers. P8 reported increased self-efficacy in Tasks 2 and 3 compared to Task 1, attributing this improvement to their growing familiarity with the process. Participants achieved these results with only a brief demonstration and minimal instruction during the 1-hour evaluation. Given the inherent complexity of the MixITS domain and the challenges faced by participants in the formative class, we consider these results promising. This indicates that MixITS-Kit effectively distills prior design experiences, guiding novice designers through the complexities of MixITS.

Our evaluation results show general participant agreement that MixITS-Kit achieved the toolkit research goals proposed by Ledo et al.~\cite{ledo2018evaluation}. We attribute these results to the combined use of the toolkit components that helped participants to tackle MixITS design problems from different levels of abstraction.
The design patterns with example solutions provided an accessible starting point, helping participants overcome the ``cold-start''~\cite{winters2009dealing} problem and begin iterating on their own solutions earlier. P6 effectively used the problem column of the catalog to find a design pattern that matched their case in Task 1. P1 highlighted the value of having standardized problems in the AI-MR domain. These observations emphasize the value of our cataloged problems, perhaps more so than the solutions, as solutions may significantly change with technology, but the core problems remain consistently relevant and future-proof.

As illustrated in Figure~\ref{fig:eval_samples}, P6 utilized descriptive AR labels (H-Ex-E-1) with instructional guidance on crashpad importance, aligning with ``Teaching and Directing.'' After considering ``Build Trust,'' P10 incorporated an AI-generated explanation for the goal inferred by AI (AI-Ev-H-36). These are examples that the high-level considerations can inspire changes in previously proposed solutions. We speculate that takin our considerations in the early stages of MixITS development can strengthen the design and avoid drastic changes later on, possibly reducing rework.

Participants 1 and 6 observed strong synergy among the toolkit's components. By approaching the design patterns with a specific consideration in mind, they successfully identified a solution. For the problem in Task 1, P6 utilized the design pattern catalog as a reference to complete the canvas. These examples illustrate how designers can benefit from the integrated synergies in MixITS-Kit, rather than relying only on the isolated use of its components.

Our results suggest that our toolkit could offer shared vocabulary among designers, as participants found its language easy to communicate with and understand. The proposed design patterns further contribute to a shared vocabulary. As suggested by Alexander et al.~\cite{alexander2018pattern}, design patterns externalize and document recurrent design decisions to foster a shared vocabulary, enabling collaborative and iterative refinement of solutions. Task 2 further supports this, with all participants successfully identifying the actor and target in the interactions.

\subsection{MixITS Class Reflection} 

The extended 10-week course, compared to a short-term workshop, provided the instructional team with deeper insights into effectively teaching MixITS system design principles. This format revealed common challenges, knowledge gaps, and misconceptions faced by novices. A key hurdle was shifting from this technology-driven mindset to a user-centered approach. Students particularly struggled with selecting appropriate interaction modalities, effective task segmentation, considering multiple task completion paths, and anticipating potential user and AI errors.

Role-play exercises with the Wizard of Oz technique allowed participants to understand user needs and technology limitations of AI and mixed reality. These test exercises revealed user difficulties such as misunderstanding the system's capabilities, performing unintended actions, attempting to interact with non-existent features, and trying to communicate with the AI as if it was another human. This didactic intervention proved effective in changing students' mindsets from technology-centered to user-centered. As students' understanding of user needs grew, they started to propose features grounded in user behavior rather than solely technological feasibility. This evolution in approach marked a significant and desirable shift that was evident in the final project designs.

Our toolkit developed from these insights and can help designers avoid common technology-centric pitfalls such as overreliance on touch interfaces, voice commands without context awareness, or assuming constant internet connectivity. Overall, the toolkit encourages designers to transcend established AI interaction paradigms, moving beyond simple prompting or screen-based interfaces towards a more holistic, embodied, and multi-modal approach essential for MixITS systems.

\section{Limitations and Future Work}

\subsection{Curricular Bias} We adopted an in-person classroom approach as a method for collecting instances of MixITS system designs and accompanying design processes. Among the limitations of this methodology choice are the particular perspectives on MixITS design arising from the choice of topics covered in class, selected reading material, the presentation approach of the class content in lectures and activities, proposed prototyping tools, and assignment requirements. Even though the teaching team made a continuous effort to foster discussions acknowledging multiple perspectives on the MixITS domain, we recognize that our teaching approach may have introduced its own biases. Moreover, we acknowledge that while the participants in our formative study and evaluation represent early-career designers, engineers, and researchers who would use MixITS, future work should expand the MixITS-Kit with data collected from more experienced professionals in the industry.

\subsection{Interactive MixITS-Kit}
Using our Toolkit, novices in the MixITS domain were able to analyze interaction problems from a user perspective, propose solutions, and refine them within an hour, including the initial onboarding. However, our evaluation revealed areas for improvement in the Canvas and a need to lower the current threshold of the toolkit. Participants reported difficulties in identifying the gulfs (P4, P5, P9) and suggested adding more detailed examples (P10) and clearer instructions (P4).

To address these issues, future work could involve developing an interactive version of MixITS-Kit, as suggested by P6. An integrated web platform that connects a browsable design pattern catalog with the interaction canvas would facilitate the use of strategies similar to those employed by P1 and P6, taking advantage of the synergy between toolkit components. This interactive version could also integrate data from user testing logs or the open-source MixITS database~\cite{bohus2024sigma} to ground designers' analyses, akin to~\cite{castelo2023argus} but with a focus on interaction design rather than modeling.

\subsection{Expanding MixITS-Kit}

Our Design Pattern catalog is intended as a starting point, and we encourage others to replicate our methodology, refine our patterns, and add new ones. Re-purposing design patterns from related areas like mixed-initiative systems and intelligent tutoring systems could grow the list of MixITS design patterns. As consumer-grade MixITS products emerge, their inclusion as case studies will also help broaden the analysis pool, as seen in other design guidelines derives from web search, activity tracking, or recommendations~\cite{amershi2019guidelines}. In this work, we refrain from including MixITS solutions we implemented in the past or hypothetical ones not proposed by students. Consequently, some patterns presented in a gulf might also be relevant in another. For example, H-Ev-E-17 (Tool Operation) could easily be applied to the gulf of execution to instruct users on how to operate a tool. We encourage designers to consider the potential for re-purposing patterns across different gulfs, interpreting them from the perspective necessary for the design of their specific MixITS systems.

Our MixITS design patterns are derived from low-fidelity prototypes. Even though these designs are plausible, the course did not focus on the engineering challenges of such systems which have been explored in prior work~\cite{bohus2021platform, anderson1985intelligent, andrist2019demonstrating, castelo2023argus}. As the realtime task guidance field evolves, implementing these patterns in productive systems and evaluating their effectiveness with real users will be crucial to validate their practicality, along the lines of an evaluation of existing systems by Amershi et al.~\cite{amershi2019guidelines}.

\section{Conclusion}
In this work, we proposed an interaction design toolkit to support the design of AI systems for intelligent task support in mixed reality (MixITS). Drawing from MixITS prototypes developed during a 10-week graduate course on human-AI interaction, we derived six design considerations, an accompanying set of 36 design patterns, and an interaction canvas to help analyze and identify gulfs of execution and evaluation between the three MixITS entities of the user, the AI, and the environment. 
Our work aims to inspire and support the design and development of MixITS systems that blend the digital and physical worlds, creating situated and context-aware learning and task guidance experiences. Successfully tackling design challenges in the MixITS domain can play a significant role in ensuring that the transformative potential of AI is accessible to a wider range of users by augmenting the process of skill acquisition.

\bibliographystyle{ACM-Reference-Format}
\bibliography{main}

\end{document}